\documentclass[prl,twocolumn,preprintnumbers,amsmath,amssymb]{revtex4-2}
\usepackage{amsmath,amssymb}        
\usepackage{siunitx}                
\usepackage{graphicx}
\usepackage{subcaption}             
\usepackage{hyperref}
\usepackage{xcolor}
\usepackage{amsmath}

\begin{document}

\title{A laser plasma soliton fusion scheme}

\author{Pisin Chen$^{a,b,c}$}
\email{pisinchen@phys.ntu.edu.tw}
\author{Yung-Kun Liu$^{a,b}$}
\email{r06222017@ntu.edu.tw}
\author{Gerard Mourou$^d$}
\email{gerard.mourou@polytechnique.edu}
\affiliation{
$^a$Leung Center for Cosmology and Particle Astrophysics, National Taiwan University, Taipei 10617, Taiwan}
\affiliation{$^b$Department of Physics, National Taiwan University, Taipei 10617, Taiwan}
\affiliation{$^c$Graduate Institute of Astrophysics, National Taiwan University, Taipei 10617, Taiwan}
\affiliation{$^d$International Center for Zetta-Exawatt Science and Technology, École Polytechnique, Palaiseau Cedex 91128, France}

\begin{abstract}
We introduce a novel and generic fusion scheme enabled by laser-plasma solitons, which promises to overcome several fundamental obstructions to reaching the breakeven condition. To demonstrate that, we invoke both deuterium-tritium (DT) and proton-boron (pB) as fuels. The intense electromagnetic field trapped inside the soliton significantly enhances the DT and pB fusion cross sections. Its ponderomotive potential evacuates electrons almost instantly, while the ions left behind are accelerated by the unshielded Coulomb field to kinetic energies suitable for fusion reaction on a longer time scale. Such a difference in time scales renders a time window for fusion to occur efficiently in an electron-free environment. We inject two consecutive lasers, where the first would excite plasma solitons and the second, much more intense and with a matched lower frequency, would fortify the soliton electromagnetic field resonantly. Solitons are known to expand during its lifetime, which significantly increases the number of ions to participate in the fusion process and the ion propagation distance. We show that the breakeven condition is attainable for both DT and pB cases. Invoking fiber laser and the iCAN laser technologies for high repetition rate and high intensity operation, gigawatt output may be conceivable.
\end{abstract}


\maketitle

 
In order to bypass the challenges of mainstream schemes to fusion, nonthermal, beam-driven, and transient approaches to fusion have been pursued \cite{Moreau1977,Rostoker1993,Rostoker1997,Eliezer2016}. However, these approaches face their own challenges. One serious issue is the shortness of the penetration depth of the fusion particles relative to their mean free path inside the target because of the smallness of the fusion cross section and the severe energy loss through collisions with the intervening electrons \cite{Li1993}. 

It happens that a laser-induced plasma soliton can address all these drawbacks in one stroke. Solitons are a fundamental phenomenon in the dynamics of laser-plasma interaction, whose property has been well characterized \cite{Kozlov:1979,Bulanov:1997,Farina:2000}. These solitons naturally occur when the laser-plasma interaction is in the non-linear regime, that is, the laser strength parameter $a_0\equiv eE/mc\omega > 1$, where $E$ is the electric field and $\omega$ is the frequency of the laser, and the plasma density slightly undercritical. Laser-plasma solitons are super-stable \cite{Hadzievskil:2006}. One salient character is that it retains a significant fraction of the laser EM field, albeit at a down-shifted frequency. 

In our scheme, we fortify the trapped laser field strength by coupling the soliton with a second laser, whose frequency matches that of the soliton. The extremely intense EM field trapped in the soliton built up through this method would significantly enhance the fusion cross section by orders of magnitude. 

The ponderomotive force associated with the trapped soliton field would evacuate electrons promptly and ions would be accelerated by the unshielded Coulomb force. The difference in the time scales between the evacuation of electrons and the motion of the ions renders a time window where the slowly moving fuel particles inside the soliton would propagate in an electron-free environment without much energy loss and a significant fraction of these fuels would fuse under the embedded ultra-intense EM field with a much enhanced cross section. 

Solitons are known to expand in size during their lifetimes \cite{bulanov_pegoraro:2002,Sarri:2010}. This effect contributes to significantly increase the number of nuclear fuels and the ion propagation distance, and is therefore one of the key ingredients in our scheme.   

The pros and cons between deuterium-tritium (DT) and proton-boron-11 ($p  ^{11}{\rm B}$, or simply $p{\rm B}$) as fusion material are well-known (see, for example, \cite{Lavell:2024}). It turns out that our scheme can overcome fundamental obstructions in both cases.  
After detailing some aspects of the underlying laser-plasma soliton dynamics, we will provide a self-consistent preliminary conceptual design for both DT and pB fusion materials, supported by 2D3V particle-in-cell (PIC) simulations performed using the EPOCH code \cite{EPOCH}, as an example to demonstrate that the breakeven condition can be reached.

\begin{figure*}[htp]
    \centering
    \includegraphics[width=1.0\linewidth]{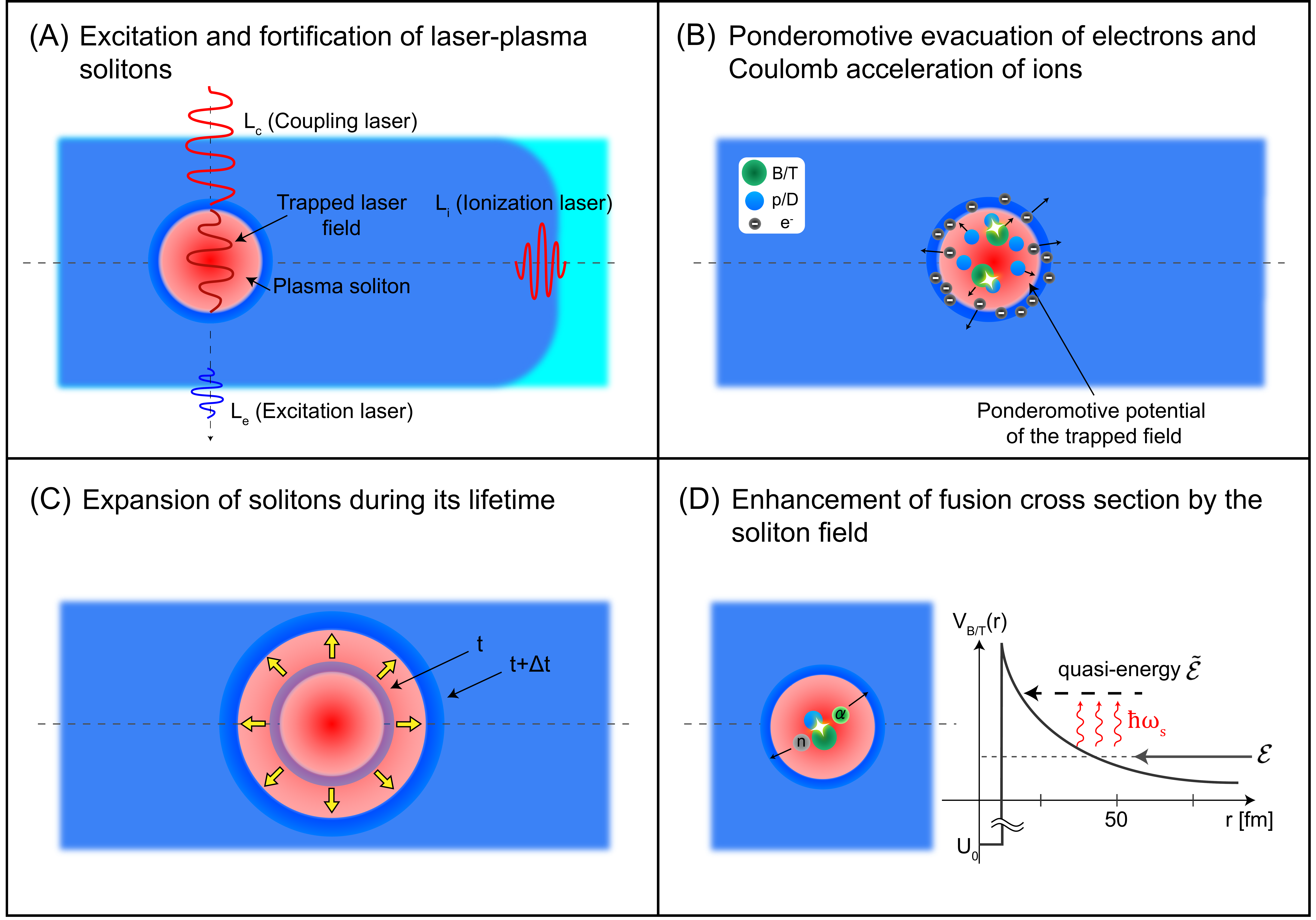} 
    \caption{Four key elements of laser-plasma soliton fusion: A. Excitation and fortification of laser-plasma solitons, B. Ponderomotive evacuation of electrons and Coulomb acceleration of ions, C. Soliton expansion during its lifetime, and D. Enhancement of the fusion cross section by the soliton field.}
    \label{fig:key_element}
\end{figure*}

\noindent\underline{\it Four key elements}

The setup of our scheme is natural and relatively simple. Specifically, our system consists of either an equal mixture of deuterium-tritium (${\rm D}_2$/${\rm T}_2$) or a diborane (${\rm B}_2{\rm H}_6$) gas target and three laser systems. First, a 200nm {\it ionization laser}, $L_i$, which is used to ionize the gas target into a plasma. Second, a UV {\it soliton excitation laser}, $L_e$, with $a_{0,e} > 1$, and the third, a long-pulse {\it coupling laser}, $L_c$, with an ultra-high intensity $a_{0,c}\equiv eA_c/m_ec=eE_c/m_ec\omega_c \gg 1$, which are injected in sequence in the transverse direction perpendicular to the axis defined by $L_i$, where $L_e$ induces a chain of $N$ solitons, followed by $L_c$ that fortifies the EM field strength in these solitons. We briefly summarize our general concept by breaking it down to four key elements (See Fig.1)

A. {\it Fortification of laser-plasma solitons} 
With the purpose of maximizing the trapped field strength of the soliton in mind, we invoke a novel mechanism to fortify it (See Appendix A for more details). The soliton created by $L_e$ traps a fraction of its energy with a frequency that is significantly down-shifted. A {\it coupling laser}, $L_c$, is then injected with a proper delay time and a frequency, $\omega_c$, that matches with this down-shifted frequency of the nascent soliton. The underlying physics is somewhat analogous to that of the resonant coupling between a EM-field and a metallic cavity \cite{Haus:1991}.

\begin{figure}[h]
\centering
\includegraphics[width=0.9\columnwidth]{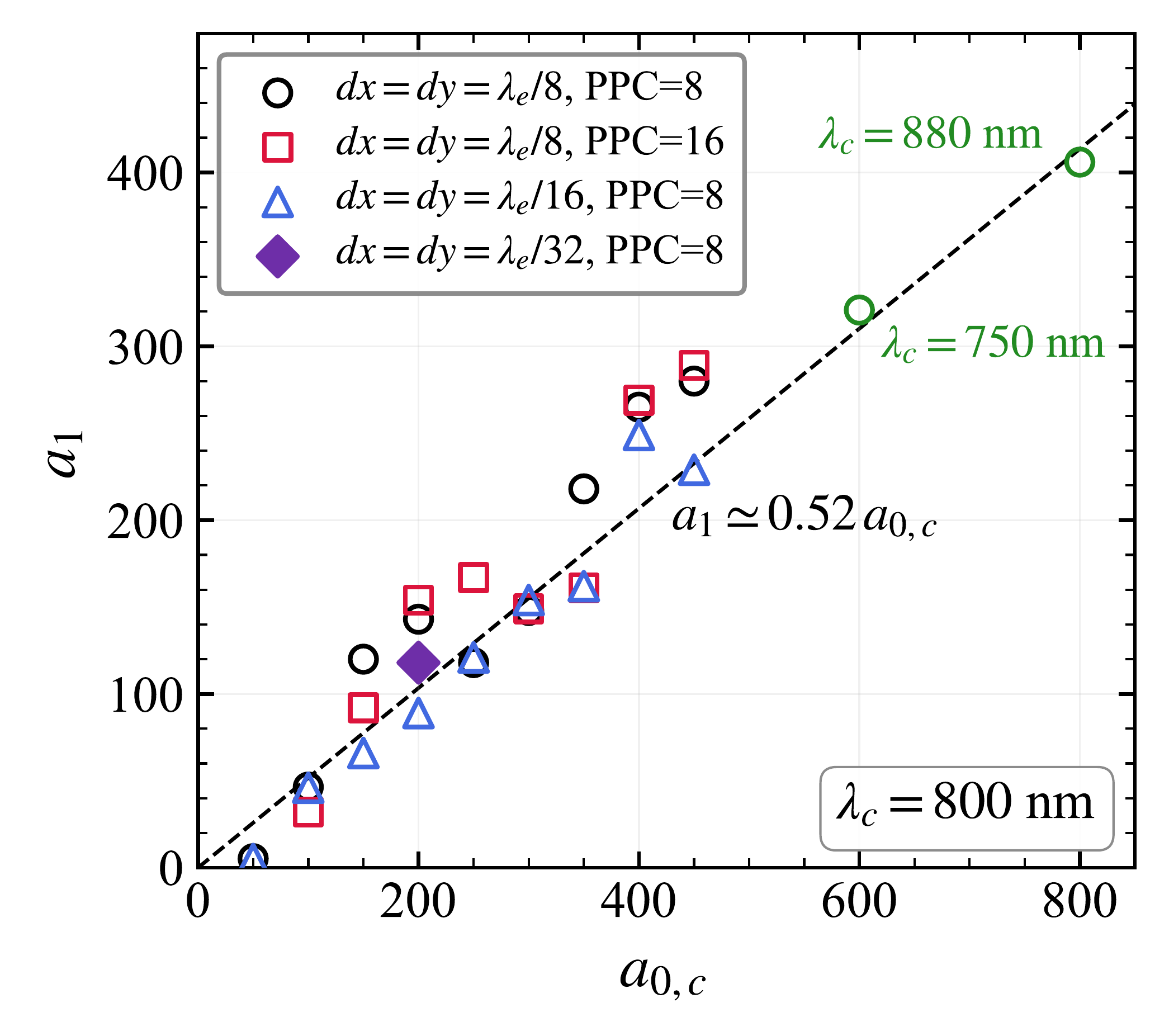}
\caption{Dependence of the fortified soliton strength parameter, $a_1$, on that of the coupling laser $a_{0,c}$. Data points in circles are the baseline results with grid size $dx=dy=\lambda_e/8$ and particles per cell PPC=8. Those in squares have the PPC doubled, and triangles have the grid size halved. 1/4 the grid size for $a_{0,c}=200$ shows (solid diamond) the convergence of our simulations.}
\label{fig:soliton_a0c_a1}
\end{figure}

Unlike the metallic cavity, soliton fortification is a highly dynamical process. During the resonant coupling, the ever-increasing ponderomotive pressure would further expand the physical size of the soliton cavity. This dynamical expansion stretches the trapped optical period, causing further frequency redshift. Consequently, the final trapped wavelength of the fully fortified soliton, $\lambda_s$, becomes significantly larger than the wavelength of the coupling laser $\lambda_c$. Due to these intricate interplays, the laser-soliton resonance tends to be not as sharp as that for metallic cavities. Let the fortified trapped laser field strength of the soliton be $a_1$. Figure~\ref{fig:soliton_a0c_a1} shows $a_1$ as a function of $a_{0,c}$ of $L_c$. We see that  under our method they scale roughly linearly as
\begin{equation}
a_1\simeq 0.52a_{0,c}, \quad\quad a_{0,c}\in[0,800].
\end{equation}

B. {\it Ponderomotive evacuation of electrons and Coulomb acceleration of ions}
As is well-known, the gradient of the EM field intensity gives rise to a ponderomotive force \cite{FChen:2015}. In the case of solitons, it pushes all the electrons outward:
\begin{align}
    F_{p}^e=-\frac{1}{4}\frac{e^2}{m_e\omega^2}\nabla E^2(r)=-\frac{1}{4}\nabla a_1^2(r) m_ec^2\sim \frac{1}{4}\frac{a_1^2}{\lambda_s}m_ec^2.
\end{align}

The ponderomotive force for ions, on the other hand, is much smaller by a factor $m_e/m_i$: $F_p^i\sim (m_e/m_i)F_p^e$. The dominant force that pushes ions outward is instead the repulsive Coulomb force among ions \cite{Naumova:2001}. In comparison, the time scales of evacuation between electrons and ions differ by a factor $\tau_e/\tau_i\sim \sqrt{m_e/m_i}$. This has been robustly validated by theoretical models \cite{Bulanov:1997, Farina:2000} and multidimensional simulations \cite{Naumova:2001,Esirkepov:2002}. 

The time window between these two scales renders the ions propagate and collide in an electron-free environment without suffering energy loss. This salient feature provides a unique environment ideal for fusion.

C. {\it Expansion of soliton during its lifetime} 
It is known that a stationary soliton would expand its size during the lifetime, which has been theoretically predicted \cite{bulanov_pegoraro:2002} and experimentally confirmed \cite{Sarri:2010}. 
The initial size of the soliton $\lambda_{s0}$ at the time $t_0$ when created would expand to $\lambda_s(\tau_s)$ at the end of its lifetime $\tau_s$. An effective soliton size, $\lambda_s$, averaged over the lifetime, gives
\begin{equation}
\lambda_s=\frac{5}{7}\Big(5\frac{\tau_s-t_0}{t_s}\Big)^{2/5}\lambda_{s0}\simeq\frac{5}{7}\Big(5\frac{\tau_s}{t_s}\Big)^{2/5}\lambda_{s0},
\end{equation}
where $t_s$ is the characteristic time scale of the expansion:
\begin{equation}
t_s=\Big[\frac{1}{2\pi}\Big(\frac{e}{q_i}\frac{m_i}{m_e}\Big)\frac{r_en_i\lambda_{s0}^4}{c^2a_1^2}\Big]^{1/2},
\end{equation}
where $m_i$ is the mass, $q_i$ the charge, and $n_i$ the density of the ion, and $r_e=e^2/m_ec^2=2.8\times 10^{-13} {\rm cm}$ is the classical electron radius. When there are more than one species of ions, $t_s$ is dominated by the heaviest ion. Our PIC simulations confirm Eq.(3) (See Appendix B). 

Note that as the soliton expands, the $a_1$ value continues to decrease. Based on the assumption of adiabatic invariance of soliton expansion \cite{Naumova:2001}, i.e., $(E_s(t)^2/\omega_s(t))\lambda_s(t)^3\sim {\rm const.}$, and since $\omega_s\propto \lambda_s(t)^{-1}$, we have $E_s(t)\propto \lambda_s(t)^{-2}$, and therefore $a_1\propto \lambda_s(t)^{-1}$.

D. {\it Enhancement of fusion cross section}
Enhancement of the fusion cross section by an ultra-intense laser field has been pursued in recent years (See \cite{Lindsey:2024,Lv:2022,Qi:2026} and references therein).

\begin{figure}
    \centering
    \includegraphics[width=1.0\columnwidth]{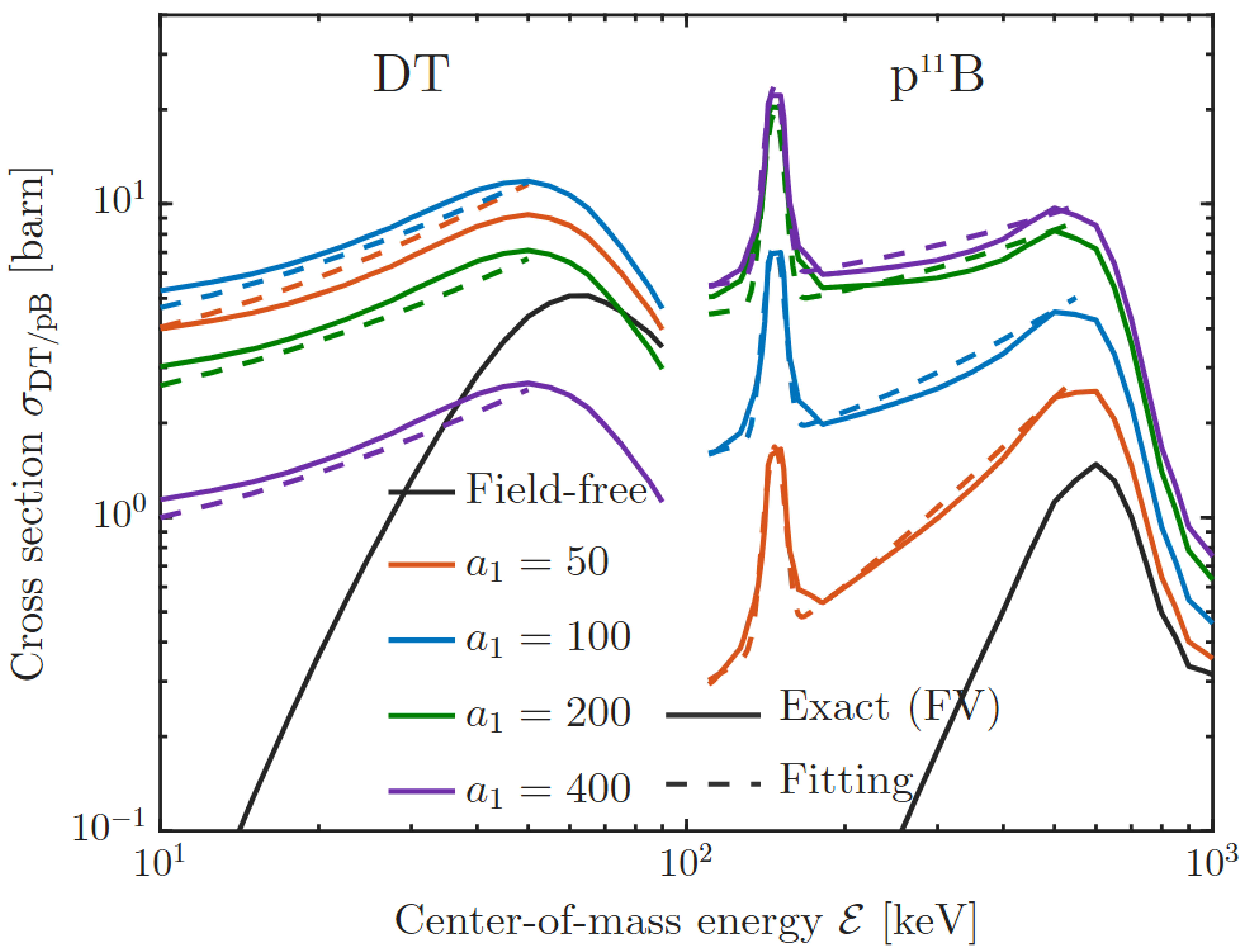}
    \caption{DT and p$^{11}$B fusion cross sections enhanced by trapped soliton field as a function of the field-free center-of-mass energy $\mathcal{E}$ of DT or pB pair. Solid curves with color denote the numerical calculations based on the FV formalism [Eq.~(\ref{eq:FV_Crosssection})], dashed curves the fitting formula [Eq.~(\ref{eq:Crosssection_fitting_formula})], and black curves the field-free cases.}
    \label{fig:pB_crosssection_different_a1}
\end{figure}

The fusion cross section is often expressed by the well-known Gamow formula,
\begin{align}
    \sigma(\mathcal{E})=\frac{1}{\mathcal{E}}S(\mathcal{E})\mathcal{T}(\mathcal{E})=\frac{1}{\mathcal{E}}S(\mathcal{E})e^{-\sqrt{\mathcal{E}_G/\mathcal{E}}},
\end{align}
where $\mathcal{E}$ is the center-of-mass energy of the fusion particles, $S(\mathcal{E})$ the astrophysical factor that governs the nuclear interaction, and $\mathcal{T}(\mathcal{E})$ the Gamow factor, which governs the tunneling probability of the repulsive Coulomb potential between the colliding ions, $\mathcal{E}_G=1.18/23.3{\rm MeV}$ is the Gamow energy for DT/pB fusion, respectively. 

The contribution of the embedded laser is through the {\it quasi-energy} between colliding ions induced by the ponderomotive energy $U_{p}=(Z_1/m_1-Z_2/m_2)^2(\mu e^2E_s^2/4\omega_s^2)$, where $\mu$ is the reduced mass, and the quivering energy $U_q\equiv \langle qAu\rangle$, where $u$ is the relative speed between ions, which results in the modification of the field-free energy to quasi-energy $\tilde{\mathcal{E}} \simeq \mathcal{E}+U_q+U_p$ \cite{Lindsey:2024} in $1/\tilde{\mathcal{E}}$ and $\mathcal{T}(\tilde{\mathcal{E}})$, while $S(\mathcal{E})$ is unaffected by the EM field.

Lindsey et al. \cite{Lindsey:2024} invoked the Floquet-Volkov (FV) formalism to obtain the laser-enhanced Gamow factor in terms of generalized Bessel functions $J_n(x,y)$:
\begin{equation}
\mathcal{T}_{\rm FV}(\mathcal{E})=\sum^{\infty}_{n=-\infty}\Big|J_n\Big(\frac{U_q}{\hbar\omega},\frac{U_p}{2\hbar\omega}\Big)\Big|^2
\mathcal{T}(\mathcal{E}_n).
\label{eq:FV_Crosssection}
\end{equation}
Here we denote the multiparameter dependence of $\mathcal{T}_{\rm FV}$ by simply $\mathcal{E}$.
They showed that this formula agrees well with the Crank-Nicolson (CN) numerical solution to the time-dependent Schr\"{o}dinger equation, while the Kramers-Henneberger (KH) model \cite{Lv:2022} leads to vastly different predictions at low $\mathcal{E}$. Given its broader applicability, we employ the FV method to calculate the enhanceed cross-section under the soliton field.

Since summing over an infinite series of generalized Bessel function is rather cumbersome, we have therefore deduced a semi-empirical fitting formula for the enhanced fusion cross section within the range of the center-of-mass energy $\mathcal{E}\in [10,50] {\rm keV}$ for DT fusion and $\mathcal{E}\in [100,500] {\rm keV}$ for pB fusion.
We find 

\begin{equation}
\sigma_{\rm DT/pB} \simeq \left\{ \begin{aligned} &\Big[6500+420\mathcal{E}\Big]\frac{e^{-\sqrt{1182/\tilde{\mathcal{E}}}}}{\tilde{\mathcal{E}}} \quad \text{(DT),}\\
    &420\Big[400+\mathcal{E}+1800f(\mathcal{E})\Big]\frac{e^{-\sqrt{22300/\tilde{\mathcal{E}}}}}{\tilde{\mathcal{E}}} \quad\text{(pB)},
\end{aligned} \right.
\label{eq:Crosssection_fitting_formula}
\end{equation}
in units of barns ($1\times 10^{-24} {\rm cm}^2$) and all energies are in units of keV. Here $f(\mathcal{E})=\exp\{-(\mathcal{E}-147)^2/64\}$ is an approximation to the pB fusion resonance peak at 147 keV, and $\tilde{\mathcal{E}}= \mathcal{E} +qa_1+pa_1^2$ is an approximation to $\tilde{\mathcal{E}} \simeq \mathcal{E}+U_q+U_p$, where $q= 3.75$ and $p = 0.036$ for pB fusion and $q=0.2$ and $p = 0.044$ for DT fusion. The square-bracket corresponds to the astrophysical factor $S(\mathcal{E})$. 

Figure~\ref{fig:pB_crosssection_different_a1} shows the laser-enhanced cross section for both DT and pB fusion as a function of the field-free center-of-mass energy $\mathcal{E}$ of DT or pB pair. The solid curves with color are the exact calculation based on the FV formalism for different values of $a_1$. The dashed curves are the fitting formula in Eq.(7). The black curves are the standard field-free cross sections.

\noindent\underline{\it Breakeven condition} 
In terms of fusion output, an intense laser can typically excite multiple solitons. We focus on the output of the first soliton only and treat that from the remaining solitons as a bonus.
The gain factor is therefore $G=\mathcal{E}_{f}/(\mathcal{E}_i+\mathcal{E}_e+\mathcal{E}_c) \geq 1$, where $\mathcal{E}_{f}$ is the output energy of fusion from the first soliton per tri-laser cycle, and the input energy is contributed from the sum of the three lasers: $L_i, L_e, L_c$. We will see later that $\mathcal{E}_i+\mathcal{E}_e\ll\mathcal{E}_c$. So we can safely ignore the first two terms in the denominator.

The energy of the coupling laser is defined as
\begin{equation}
\mathcal{E}_c=\frac{E_c^2}{4\pi}V_c=\frac{\pi a_{0,c}^2 m_ec^2}{r_e \lambda_c^2}\left(2\pi r_c^2\tau_cc\right)= 2\pi^2a_{0,c}^2\frac{\tau_cc}{r_e}m_ec^2,
\end{equation}
where $V_c$ is the volume and $a_{0,c}$ is the strength parameter of the coupling laser, $\lambda_c$ is its wavelength, $r_c$ the transverse size, $\tau_c$ the pulse duration. Here we choose $r_c\simeq \lambda_c$ so that the transverse sizes of $L_e$ and $L_c$ are roughly matched. 

On the other hand, the fusion output energy is
\begin{equation}
\mathcal{E}_{f}=N_{\rm D/p}P_{\rm DT/pB}\epsilon_{\rm DT/pB},
\end{equation}
where $P_{\rm DT/pB}$ is the probability of fusion per tritium or boron in the soliton and $\epsilon_{DT/pB}=17.6/8.7{\rm MeV}$ is the energy released per DT/pB fusion. $N_{\rm D/p}$ is the total number of deuterium/proton nuclei within the effective volume $V_s$ of the soliton during its lifetime, that is,
\begin{equation}
N_{\rm D/p}=n_{\rm D/p}V_s=n_{\rm D/p}(\sqrt{2\pi} \lambda_s)^3. 
\end{equation}
The density of proton and boron nuclei in a completely ionized diborane gas, ${\rm B}_2{\rm H}_6$, are $n_p=3/8 n_p$ and $n_B=1/8 n_p$, where $n_p$ is the electron (and therefore the plasma) density, while that for an equally mixed ${\rm D}_2/{\rm T}_2$ gas is $n_D=n_T=1/2 n_p$.
The total probability of DT/pB fusion that occurs inside the electron-free soliton, is 
\begin{equation}
P_{\rm pB}
= \sigma_{\rm DT/pB}\frac{n_{\rm T/B}V_s}{2\pi\lambda_s^2}=\sqrt{2\pi}\sigma_{\rm DT/pB}n_{\rm T/B}\lambda_s,
\end{equation}
where $\lambda_s$ is the time-averaged  wavelength of the trapped laser field as well as the size of the soliton.

Putting all these together, we obtain the gain factor in terms of $a_1$:
\begin{equation}
G\simeq \frac{1.85}{a_1^2}\frac{r_e}{\tau_cc}\Big(\frac{\tau_s}{t_s}\Big)^{8/5}n_{\rm D/p}n_{\rm T/B}\lambda_{s0}^4\sigma_{\rm DT/pB}\frac{\epsilon_{\rm DT/pB}}{m_ec^2}.
\end{equation}
As commented, $a_1$ is not a constant but decreases during its lifetime. However, it can be verified that due the the counter-balance among the several $a_1$-dependent parameters, $G$ is insensitive to the time-dependence of $a_1$. 

\noindent\underline{\it A preliminary conceptual design} 

Based on the considerations above, here we provide a preliminary conceptual design of our fusion scheme with two numerical examples for pB and DT fusion.
PIC simulations indicate that the excitation of solitons and trapping of the laser field are optimal when the ratio of plasma density to laser critical density, $\phi_0\equiv n_p/n_c$, is $\sim 0.1$, where the critical density is defined as $n_c=\pi/(r_e\lambda_s^2)$. 

The configuration of the tri-laser system follows the description in Fig.1, where the orientation of the ionization laser is perpendicular to those of the excitation and the coupling lasers. The coupling laser should follow behind the excitation laser with a proper timing to achieve the optimal condition for the fortification of the soliton field.

We choose 750-800nm Ti:Sapphire laser for the coupling laser. PIC simulations show that a ratio of $\lambda_c/\lambda_e\sim {\mathcal{O}}(10)$ works the best for the resonant coupling (Appendix A), which determines the choice of frequency for the excitation laser with $\lambda_e=72{\rm nm}$. The plasma (electron) density of $n_p=2\times 10^{22}{\rm cm}^{-3}$ is then determined. 

Finally, we choose $a_{0,c}=200/600$ for DT/pB, respectively, so that the corresponding $a_1\sim 100/300$ would maximize $G$. These in turn determine additional parameters of $L_c$. Key parameters for DT and pB fusions are given in Table I. PIC simulations indicate that solitons created under such plasma density has an initial size $\lambda_{s0}\sim 3.5/6 \mu{\rm m}$ and a lifetime $\tau_s\sim 60 {\rm psec}$ (See Appendix B). During its lifetime, the soliton size would expand more than 10 times according to Eq.(3) and Eq.(4).

\begin{table}[htbp] 
\centering 
\begin{tabular}{|c|c|c|c|c|c|}
\hline
\multicolumn{6}{|c|}{\textbf{Case I: Deuterium-Tritium (DT) Fusion}} \\ \hline
\multicolumn{6}{|c|}{Laser parameters} \\ \hline
      & $\lambda_i$  & $a_{0,i}$ & Spot size    & Pulse duration   & Input Energy \\ \hline
$L_i$ & $200\,\mathrm{nm}$ & 1    & $20\lambda_i$ & $20\lambda_i/c$ & $0.9\,\mathrm{J}$       \\ \hline
$L_e$ & $72\,\mathrm{nm}$   & 8    & $5\lambda_s$ & $5\lambda_s/c$  & $0.3\,\mathrm{J}$      \\ \hline
$L_c$ & $800\,\mathrm{nm}$    & 200  & $1\lambda_c$ & $10\lambda_c/c$  & $179\,\mathrm{J}$    \\ \hline
\multicolumn{6}{|c|}{Soliton parameters} \\ \hline
      & $\lambda_{s0}$  & $a_1$      & $\tau_s$         & $\mathcal{E}_{CM}$& Output Energy \\ \hline
      & $3.5\,\mu\mathrm{m}$ & 100  & $60\,\mathrm{ps}$& $50\,\mathrm{keV}$ &$408\,\mathrm{J} (390\,\mathrm{J})$ \\ \hline

\multicolumn{6}{|c|}{\textbf{Case II: Proton-Boron (pB) Fusion}} \\ \hline
\multicolumn{6}{|c|}{Laser parameters} \\ \hline
      & $\lambda_i$  & $a_{0,i}$ & Spot size    & Pulse duration   & Input Energy \\ \hline
$L_i$ & $200\,\mathrm{nm}$ & 1    & $20\lambda_i$ & $20\lambda_i/c$ & $0.9\,\mathrm{J}$       \\ \hline
$L_e$ & $72\,\mathrm{nm}$   & 8    & $5\lambda_s$ & $5\lambda_s/c$  & $0.3\,\mathrm{J}$      \\ \hline
$L_c$ & $750\,\mathrm{nm}$    & 600  & $1\lambda_c$ & $10\lambda_c/c$  & $1513\,\mathrm{J}$    \\ \hline
\multicolumn{6}{|c|}{Soliton parameters} \\ \hline
      & $\lambda_{s0}$  & $a_1$     & $\tau_s$         & $\mathcal{E}_{CM}$& Output Energy \\ \hline
      & $6\,\mu\mathrm{m}$ & 300  & $60\,\mathrm{ps}$& $150\,\mathrm{keV}$ &$2289\,\mathrm{J} (2212\,\mathrm{J})$ \\ \hline
\end{tabular}
\caption{Preliminary conceptual design parameters for DT and pB fusion schemes at a background plasma density of $n_p=2\times 10^{22}\,\mathrm{cm}^{-3}$. The output energies are computed numerically using the FV formula, and that in the parentheses use the fitting formula.}
\label{table-laser_parameters}
\end{table}

The gain factor based on Eq.(12) is therefore
\begin{equation} G\simeq\frac{\mathcal{E}_{f}}{\mathcal{E}_c}\simeq 
\left\{ \begin{aligned} & 2.28 \quad \text{(DT)}\\
    & 1.51 \quad \text{(pB)}
\end{aligned} \right.,
 \label{eq:DT_gain_design}
\end{equation}
which shows that the breakeven condition is in principle attainable for both DT and pB fusion in our scheme.

We emphasize that this preliminary conceptual design is not optimized. For example, instead of a fixed frequency, if the frequency of the coupling laser is anti-chirped, or if it is split into multiple pulses with successive decreasing frequencies that are commensurate with the gradual down-shift of the soliton frequency, then perhaps the energy transfer would be more efficient.

\underline{Laser technology} 
We choose the wavelength of coupling laser $L_c$, $\lambda_c=750-800{\rm nm}$, for obvious reason since the technology for ultra-intense lasers at this wavelength is already mature, although the typical energy per pulse in state-of-the-art lasers is still substantially below those listed in Table I. The $L_e$ laser with wavelength at $\lambda_e=72{\rm nm}\sim \lambda_c/10$ may be prepared by frequency double-tripling from a 700 nm laser. Although the efficiency would be compromised in so doing, fortunately, the pulse energy required for $L_e$ is minute in comparison with that for $L_c$. 

We note that fiber lasers are already capable of delivering MHz repetition rates \cite{Chowdhury:2019} in routine operation, albeit with low peak power. 
On the other hand, the ICAN (International Coherent Amplification Network) consortium \cite{Mourou:2013}, which invokes fiber lasers in a coherent manner, has been aiming to develop laser systems that would deliver $>10 {\rm J} $ energy per pulse, with $>10 {\rm kHz}$ repetition rate, and 100–200 femtosecond pulse duration. It is hoped that fiber-laser-based high-intensity, high-repetition-rate lasers will be developed in the near future. By then this laser-plasma soliton fusion scheme would become closer to practical applications.

We thank Donna Strickland, Nicholas Yi-Huan Chen, Ching-En Lin and Polly Ting-An Wang for helpful suggestions and comments. We thank the National Center for High-performance Computing (NCHC) in Taiwan for providing computational and storage resources. This work is supported by Leung Center for Cosmology and Particle Astrophysics (LeCosPA), National Taiwan University.
\appendix

\section{Resonant fortification of soliton field}

In our conception, the excitation laser $L_e$ ($\lambda_e = 72\,\mathrm{nm}$, $a_{0,e} = 8$, FWHM $1.2\,$fs, spot radius $r_e = 5\lambda_e$) plays a dual role: it excites the soliton and it ponderomotively creates a lower-density channel along its path in the plasma. A $20\,$fs delay allows sufficient time for channel formation for the coupling laser $L_c$ (duration $10\lambda_c/c$, $r_c = \lambda_c$) to penetrate the plasma and reach the soliton to resonantly pump the cavity. During fortification, the energy injected by $L_c$ exerts extreme ponderomotive pressure, driving a dynamical expansion of the cavity and a corresponding frequency redshift of the trapped field. 

\begin{figure}[htbp]
\centering
\includegraphics[width=0.9\columnwidth]{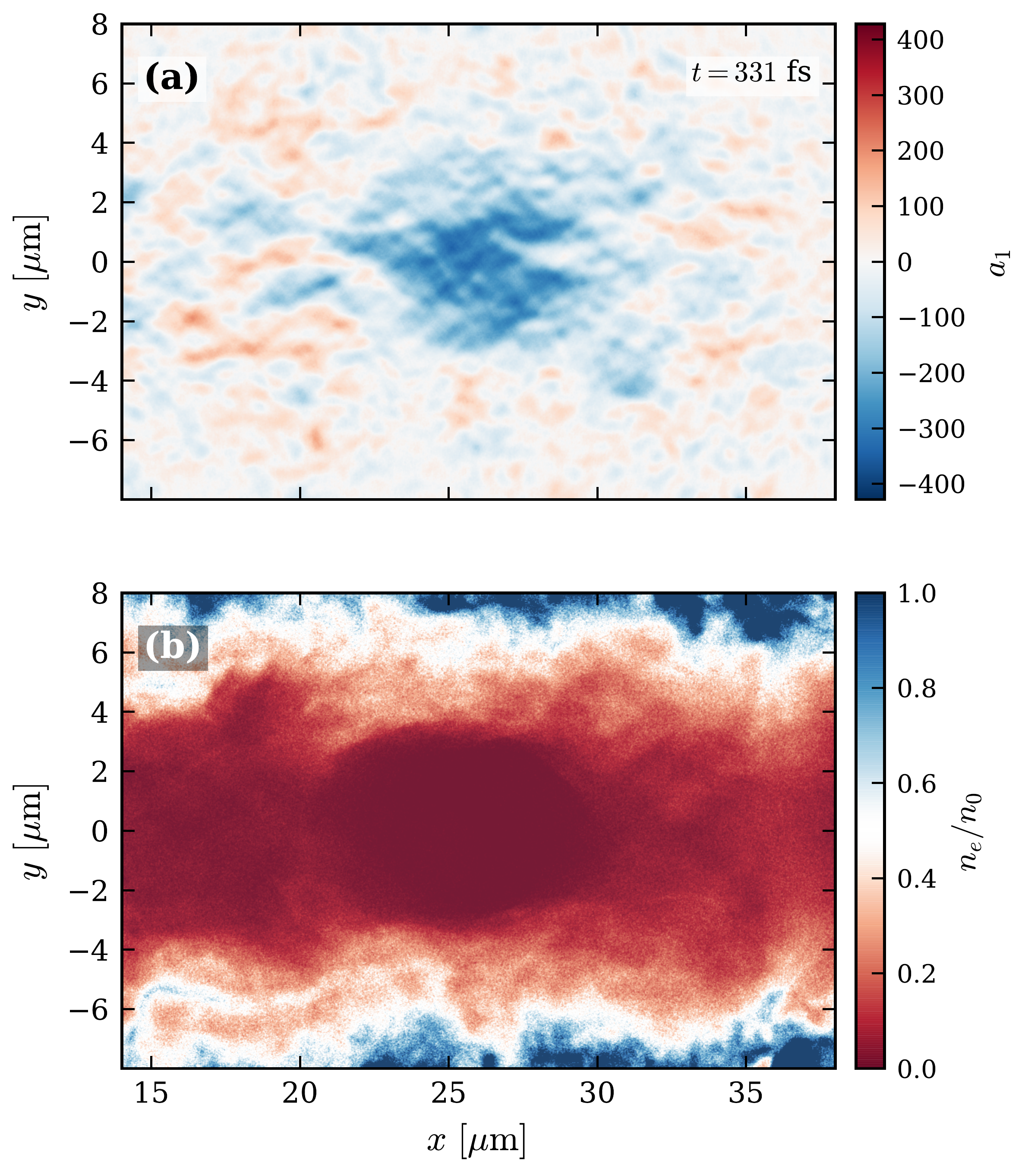}
\caption{ Snapshot of the fortified laser-plasma soliton from a 2D PIC simulation at $t=331\,$fs. The spatial distribution of the trapped EM field confirms the formation of the soliton cavity, with the peak amplitude reaching $a_1 \sim 400$.}
\label{fig:soliton_snapshot}
\end{figure}

To validate this mechanism and to extract the scaling shown in Fig.~\ref{fig:soliton_a0c_a1}, we performed 2D particle-in-cell (PIC) simulations using the EPOCH code\cite{EPOCH}. The baseline domain is $28.8 \times 20\,\mu\mathrm{m}^2$ with a $9\,\mathrm{nm}$ resolution and 8 particles per cell. Convergence tests, including doubling the spatial resolution to $4.5\,\mathrm{nm}$ and particles per cell to 16, were conducted. These tests confirm the numerical stability of the overall amplification trend. Figure \ref{fig:soliton_snapshot} presents a snapshot from one of our 2D PIC simulations with $a_{0,c}=800$, which visually demonstrates the formation of a fortified soliton with a peak trapped field reaching $a_1 \sim 400$.

\begin{figure}[htbp]
\centering
\includegraphics[width=0.9\columnwidth]{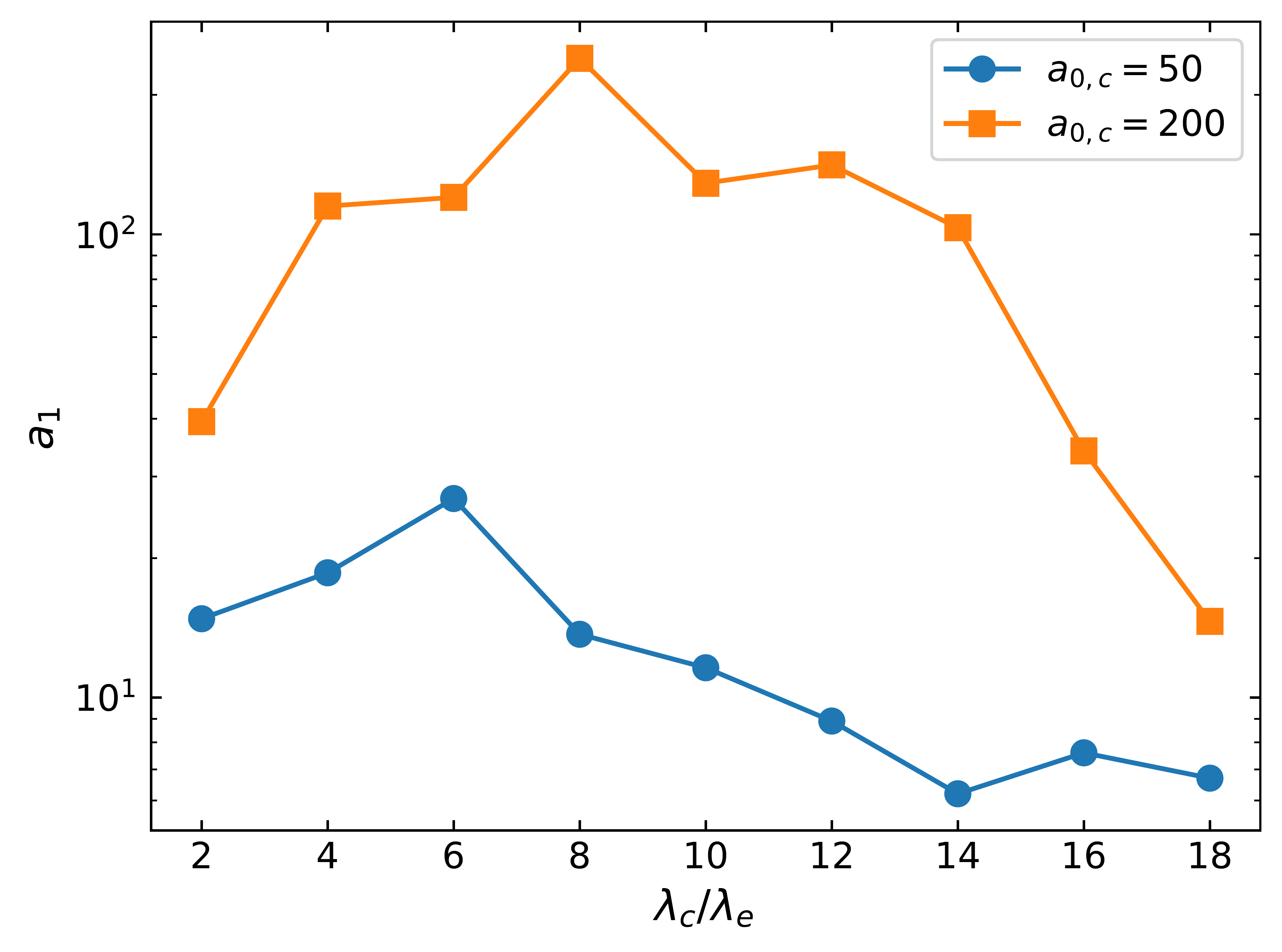}
\caption{ Evidence of resonant coupling in fortifying soliton field strength $a_1$ by scanning through the wavelength ratio $\lambda_c/\lambda_e$ for two different coupling laser intensities ($a_{0,c} = 50$ and $200$), performed at a plasma density of $n_e=10^{20}\,\mathrm{cm}^{-3}$}
\label{fig:soliton_resonance}
\end{figure}

To illustrate the resonant nature of this fortification, Fig.~\ref{fig:soliton_resonance} exhibits the dependence of the amplified field $a_1$ on the coupling laser wavelength $\lambda_c$. While this specific parameter scan was performed at a representative background density of $n_e = 10^{20}\,\mathrm{cm}^{-3}$ so as to map the fundamental trend efficiently, it clearly reveals the existence of an optimal wavelength ratio that maximizes the energy transfer. 

The physics governing this resonant amplification is analogous to that for metallic cavities. However, unlike metallic cavities where the size is fixed, a soliton under tremendous ponderomotive pressure exerted by the coupling laser would expand, and the trapped EM field wavelength in the soliton increases correspondingly. As a result, the resonance of the laser-soliton coupling tends not to exhibit as sharp a peak.  

\section{Soliton lifetime}

The lifetime of a laser-plasma soliton is fundamentally governed by the ion dynamics. Once ions respond to the unshielded Coulomb field, the structure transitions into an expanding post-soliton \cite{Naumova:2001} stage, where its evolution is best characterized by the inverse of the ion plasma frequency, $\omega_{pi}^{-1}$. Experimentally, Sarri \textit{et al.} \cite{Sarri:2010} observed that post-solitons persist for at least $140\,\mathrm{ps}$ in a deuterium plasma with density $n_e \sim 10^{20}\,\mathrm{cm}^{-3}$. At this density, the observation window corresponds to $> 1300\,\omega_{pi}^{-1}$. Note that this represents a strict lower bound, since in their experiment the solitons exhibited no sign of dissipation when their optical probe delay-line reached the technical limit.

In our preliminary design ($n_e = 2 \times 10^{22}\,\mathrm{cm}^{-3}$), the assumed soliton lifetime of $\tau_s = 60\,\mathrm{ps}$ corresponds to roughly $\sim 7100\,\omega_{pi}^{-1}$, which is roughly a factor six longer than the established experimental lower bound described above. This, however, is a highly plausible estimate that can be justified by two stabilizing mechanisms unique to our ultra-intensity regime of soliton. 

\begin{figure}[htbp]
\centering
\includegraphics[width=0.9\columnwidth]{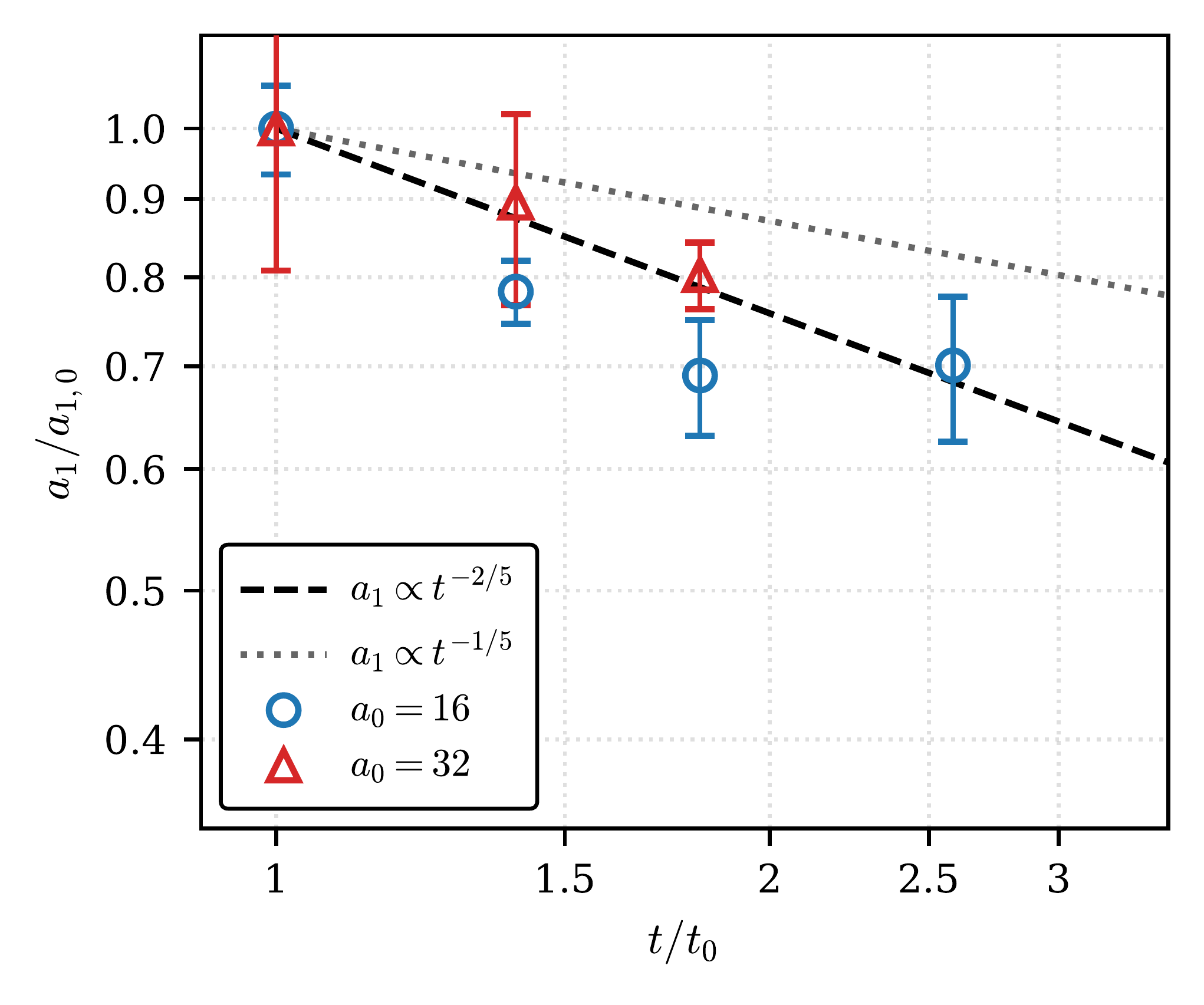}
\caption{ Temporal evolution of the normalized trapped field amplitude $a_1/a_{1,0}$, with $t_0=350\,$fs. Data extracted from 2D PIC simulations ($a_0 = 16$, blue circles; $a_0 = 32$, red triangles) trace individual solitons over time. The results consistently follow the adiabatic-invariant scaling $a_1 \propto t^{-2/5}$ (solid black line) \cite{Naumova:2001}, confirming a stable cavity expansion without sudden energy dissipation.}
\label{fig:soliton_lifetime}
\end{figure}

As shown in Fig.~\ref{fig:soliton_lifetime}, we performed 2D PIC simulations tracking the temporal evolution of $a_1$ under different initial values ($a_1(0) = 16$ and $32$). The decay of the normalized peak amplitude aligns with the theoretical prediction based on the assumption of adiabatic-invariance: $a_1 \propto t^{-2/5}$ by Naumova \textit{et al.} \cite{Naumova:2001}. This confirms that the soliton remains fully intact and undergoes a stable macroscopic expansion driven by ion momentum, completely devoid of rapid electromagnetic leakage or catastrophic hydrodynamic instabilities. Since the post-soliton evolves continuously and predictably along the adiabatic trajectory, it is safe to extrapolate the time-dependence of $a_1(t)$ to $\tau_s = 60\,\mathrm{ps}$.

\bibliography{reference}

\begin{thebibliography}{22}%
\makeatletter
\providecommand \@ifxundefined [1]{%
 \@ifx{#1\undefined}
}%
\providecommand \@ifnum [1]{%
 \ifnum #1\expandafter \@firstoftwo
 \else \expandafter \@secondoftwo
 \fi
}%
\providecommand \@ifx [1]{%
 \ifx #1\expandafter \@firstoftwo
 \else \expandafter \@secondoftwo
 \fi
}%
\providecommand \natexlab [1]{#1}%
\providecommand \enquote  [1]{``#1''}%
\providecommand \bibnamefont  [1]{#1}%
\providecommand \bibfnamefont [1]{#1}%
\providecommand \citenamefont [1]{#1}%
\providecommand \href@noop [0]{\@secondoftwo}%
\providecommand \href [0]{\begingroup \@sanitize@url \@href}%
\providecommand \@href[1]{\@@startlink{#1}\@@href}%
\providecommand \@@href[1]{\endgroup#1\@@endlink}%
\providecommand \@sanitize@url [0]{\catcode `\\12\catcode `\$12\catcode `\&12\catcode `\#12\catcode `\^12\catcode `\_12\catcode `\%12\relax}%
\providecommand \@@startlink[1]{}%
\providecommand \@@endlink[0]{}%
\providecommand \url  [0]{\begingroup\@sanitize@url \@url }%
\providecommand \@url [1]{\endgroup\@href {#1}{\urlprefix }}%
\providecommand \urlprefix  [0]{URL }%
\providecommand \Eprint [0]{\href }%
\providecommand \doibase [0]{https://doi.org/}%
\providecommand \selectlanguage [0]{\@gobble}%
\providecommand \bibinfo  [0]{\@secondoftwo}%
\providecommand \bibfield  [0]{\@secondoftwo}%
\providecommand \translation [1]{[#1]}%
\providecommand \BibitemOpen [0]{}%
\providecommand \bibitemStop [0]{}%
\providecommand \bibitemNoStop [0]{.\EOS\space}%
\providecommand \EOS [0]{\spacefactor3000\relax}%
\providecommand \BibitemShut  [1]{\csname bibitem#1\endcsname}%
\let\auto@bib@innerbib\@empty
\bibitem [{\citenamefont {Moreau}(1977)}]{Moreau1977}%
  \BibitemOpen
  \bibfield  {author} {\bibinfo {author} {\bibfnamefont {D.~C.}\ \bibnamefont {Moreau}},\ }\bibfield  {title} {\bibinfo {title} {Potentiality of the proton-boron fuel for controlled thermonuclear fusion},\ }\href@noop {} {\bibfield  {journal} {\bibinfo  {journal} {Nuclear Fusion}\ }\textbf {\bibinfo {volume} {17}},\ \bibinfo {pages} {13} (\bibinfo {year} {1977})}\BibitemShut {NoStop}%
\bibitem [{\citenamefont {Rostoker}\ \emph {et~al.}(1993)\citenamefont {Rostoker}, \citenamefont {Wessel}, \citenamefont {Rahman}, \citenamefont {Maglich}, \citenamefont {Spivey},\ and\ \citenamefont {Fisher}}]{Rostoker1993}%
  \BibitemOpen
  \bibfield  {author} {\bibinfo {author} {\bibfnamefont {N.}~\bibnamefont {Rostoker}}, \bibinfo {author} {\bibfnamefont {F.}~\bibnamefont {Wessel}}, \bibinfo {author} {\bibfnamefont {H.}~\bibnamefont {Rahman}}, \bibinfo {author} {\bibfnamefont {B.~C.}\ \bibnamefont {Maglich}}, \bibinfo {author} {\bibfnamefont {B.}~\bibnamefont {Spivey}},\ and\ \bibinfo {author} {\bibfnamefont {A.}~\bibnamefont {Fisher}},\ }\bibfield  {title} {\bibinfo {title} {Magnetic fusion with high energy self-colliding ion beams},\ }\href {https://doi.org/10.1103/PhysRevLett.70.1818} {\bibfield  {journal} {\bibinfo  {journal} {Phys. Rev. Lett.}\ }\textbf {\bibinfo {volume} {70}},\ \bibinfo {pages} {1818} (\bibinfo {year} {1993})}\BibitemShut {NoStop}%
\bibitem [{\citenamefont {Rostoker}\ \emph {et~al.}(1997)\citenamefont {Rostoker}, \citenamefont {Binderbauer},\ and\ \citenamefont {Monkhorst}}]{Rostoker1997}%
  \BibitemOpen
  \bibfield  {author} {\bibinfo {author} {\bibfnamefont {N.}~\bibnamefont {Rostoker}}, \bibinfo {author} {\bibfnamefont {M.~W.}\ \bibnamefont {Binderbauer}},\ and\ \bibinfo {author} {\bibfnamefont {H.~J.}\ \bibnamefont {Monkhorst}},\ }\bibfield  {title} {\bibinfo {title} {Colliding beam fusion reactor},\ }\href@noop {} {\bibfield  {journal} {\bibinfo  {journal} {Science}\ }\textbf {\bibinfo {volume} {278}},\ \bibinfo {pages} {1419} (\bibinfo {year} {1997})}\BibitemShut {NoStop}%
\bibitem [{\citenamefont {Eliezer}\ \emph {et~al.}(2016)\citenamefont {Eliezer}, \citenamefont {Hora}, \citenamefont {Korn}, \citenamefont {Nissim},\ and\ \citenamefont {Martinez~Val}}]{Eliezer2016}%
  \BibitemOpen
  \bibfield  {author} {\bibinfo {author} {\bibfnamefont {S.}~\bibnamefont {Eliezer}}, \bibinfo {author} {\bibfnamefont {H.}~\bibnamefont {Hora}}, \bibinfo {author} {\bibfnamefont {G.}~\bibnamefont {Korn}}, \bibinfo {author} {\bibfnamefont {N.}~\bibnamefont {Nissim}},\ and\ \bibinfo {author} {\bibfnamefont {J.~M.}\ \bibnamefont {Martinez~Val}},\ }\bibfield  {title} {\bibinfo {title} {Avalanche proton-boron fusion based on elastic nuclear collisions},\ }\href@noop {} {\bibfield  {journal} {\bibinfo  {journal} {Physics of Plasmas}\ }\textbf {\bibinfo {volume} {23}} (\bibinfo {year} {2016})}\BibitemShut {NoStop}%
\bibitem [{\citenamefont {Li}\ and\ \citenamefont {Petrasso}(1993)}]{Li1993}%
  \BibitemOpen
  \bibfield  {author} {\bibinfo {author} {\bibfnamefont {C.-K.}\ \bibnamefont {Li}}\ and\ \bibinfo {author} {\bibfnamefont {R.~D.}\ \bibnamefont {Petrasso}},\ }\bibfield  {title} {\bibinfo {title} {Charged-particle stopping powers in inertial confinement fusion plasmas},\ }\href@noop {} {\bibfield  {journal} {\bibinfo  {journal} {Physical review letters}\ }\textbf {\bibinfo {volume} {70}},\ \bibinfo {pages} {3059} (\bibinfo {year} {1993})}\BibitemShut {NoStop}%
\bibitem [{\citenamefont {Kozlov}\ \emph {et~al.}(1979)\citenamefont {Kozlov}, \citenamefont {Litvak},\ and\ \citenamefont {Suvorov}}]{Kozlov:1979}%
  \BibitemOpen
  \bibfield  {author} {\bibinfo {author} {\bibfnamefont {V.}~\bibnamefont {Kozlov}}, \bibinfo {author} {\bibfnamefont {A.}~\bibnamefont {Litvak}},\ and\ \bibinfo {author} {\bibfnamefont {E.}~\bibnamefont {Suvorov}},\ }\bibfield  {title} {\bibinfo {title} {Envelope solitons of relativistic strong electromagnetic waves},\ }\href@noop {} {\bibfield  {journal} {\bibinfo  {journal} {Sov. Phys. JETP}\ }\textbf {\bibinfo {volume} {49}},\ \bibinfo {pages} {75} (\bibinfo {year} {1979})}\BibitemShut {NoStop}%
\bibitem [{\citenamefont {Bulanov}\ \emph {et~al.}(1997)\citenamefont {Bulanov}, \citenamefont {Lontano},\ and\ \citenamefont {Sasorov}}]{Bulanov:1997}%
  \BibitemOpen
  \bibfield  {author} {\bibinfo {author} {\bibfnamefont {S.}~\bibnamefont {Bulanov}}, \bibinfo {author} {\bibfnamefont {M.}~\bibnamefont {Lontano}},\ and\ \bibinfo {author} {\bibfnamefont {P.}~\bibnamefont {Sasorov}},\ }\bibfield  {title} {\bibinfo {title} {Ionization rate in the presence of runaway electrons},\ }\href@noop {} {\bibfield  {journal} {\bibinfo  {journal} {Physics of Plasmas}\ }\textbf {\bibinfo {volume} {4}},\ \bibinfo {pages} {931} (\bibinfo {year} {1997})}\BibitemShut {NoStop}%
\bibitem [{\citenamefont {Farina}\ and\ \citenamefont {Bulanov}(2001)}]{Farina:2000}%
  \BibitemOpen
  \bibfield  {author} {\bibinfo {author} {\bibfnamefont {D.}~\bibnamefont {Farina}}\ and\ \bibinfo {author} {\bibfnamefont {S.}~\bibnamefont {Bulanov}},\ }\bibfield  {title} {\bibinfo {title} {Slow electromagnetic solitons in electron-ion plasmas},\ }\href@noop {} {\bibfield  {journal} {\bibinfo  {journal} {Plasma Physics Reports}\ }\textbf {\bibinfo {volume} {27}},\ \bibinfo {pages} {641} (\bibinfo {year} {2001})}\BibitemShut {NoStop}%
\bibitem [{\citenamefont {Hadzievski}\ \emph {et~al.}(2007)\citenamefont {Hadzievski}, \citenamefont {Mancic},\ and\ \citenamefont {Skoric}}]{Hadzievskil:2006}%
  \BibitemOpen
  \bibfield  {author} {\bibinfo {author} {\bibfnamefont {L.}~\bibnamefont {Hadzievski}}, \bibinfo {author} {\bibfnamefont {A.}~\bibnamefont {Mancic}},\ and\ \bibinfo {author} {\bibfnamefont {M.}~\bibnamefont {Skoric}},\ }\bibfield  {title} {\bibinfo {title} {Dynamics of weakly relativistic electromagnetic solitons in laser plasmas},\ }\href@noop {} {\bibfield  {journal} {\bibinfo  {journal} {Publications of the Astronomical Observatory of Belgrade}\ }\textbf {\bibinfo {volume} {82}},\ \bibinfo {pages} {101} (\bibinfo {year} {2007})}\BibitemShut {NoStop}%
\bibitem [{\citenamefont {Bulanov}\ and\ \citenamefont {Pegoraro}(2002)}]{bulanov_pegoraro:2002}%
  \BibitemOpen
  \bibfield  {author} {\bibinfo {author} {\bibfnamefont {S.}~\bibnamefont {Bulanov}}\ and\ \bibinfo {author} {\bibfnamefont {F.}~\bibnamefont {Pegoraro}},\ }\bibfield  {title} {\bibinfo {title} {Stability of a mass accreting shell expanding in a plasma},\ }\href@noop {} {\bibfield  {journal} {\bibinfo  {journal} {Physical Review E}\ }\textbf {\bibinfo {volume} {65}},\ \bibinfo {pages} {066405} (\bibinfo {year} {2002})}\BibitemShut {NoStop}%
\bibitem [{\citenamefont {Sarri}\ \emph {et~al.}(2010)\citenamefont {Sarri}, \citenamefont {Singh}, \citenamefont {Davies}, \citenamefont {Fiuza}, \citenamefont {Lancaster}, \citenamefont {Clark}, \citenamefont {Hassan}, \citenamefont {Jiang}, \citenamefont {Kageiwa}, \citenamefont {Lopes} \emph {et~al.}}]{Sarri:2010}%
  \BibitemOpen
  \bibfield  {author} {\bibinfo {author} {\bibfnamefont {G.}~\bibnamefont {Sarri}}, \bibinfo {author} {\bibfnamefont {D.}~\bibnamefont {Singh}}, \bibinfo {author} {\bibfnamefont {J.}~\bibnamefont {Davies}}, \bibinfo {author} {\bibfnamefont {F.}~\bibnamefont {Fiuza}}, \bibinfo {author} {\bibfnamefont {K.}~\bibnamefont {Lancaster}}, \bibinfo {author} {\bibfnamefont {E.}~\bibnamefont {Clark}}, \bibinfo {author} {\bibfnamefont {S.}~\bibnamefont {Hassan}}, \bibinfo {author} {\bibfnamefont {J.}~\bibnamefont {Jiang}}, \bibinfo {author} {\bibfnamefont {N.}~\bibnamefont {Kageiwa}}, \bibinfo {author} {\bibfnamefont {N.}~\bibnamefont {Lopes}}, \emph {et~al.},\ }\bibfield  {title} {\bibinfo {title} {Observation of postsoliton expansion following laser propagation through an underdense plasma},\ }\href@noop {} {\bibfield  {journal} {\bibinfo  {journal} {Physical review letters}\ }\textbf {\bibinfo {volume} {105}},\ \bibinfo {pages} {175007} (\bibinfo {year} {2010})}\BibitemShut {NoStop}%
\bibitem [{\citenamefont {Lavell}\ \emph {et~al.}(2024)\citenamefont {Lavell}, \citenamefont {Kish}, \citenamefont {Sexton}, \citenamefont {Evans}, \citenamefont {Mohammad}, \citenamefont {Gomez-Ramirez}, \citenamefont {Scullin}, \citenamefont {Borscz}, \citenamefont {Pikuz}, \citenamefont {Mehlhorn} \emph {et~al.}}]{Lavell:2024}%
  \BibitemOpen
  \bibfield  {author} {\bibinfo {author} {\bibfnamefont {M.~J.}\ \bibnamefont {Lavell}}, \bibinfo {author} {\bibfnamefont {A.~J.}\ \bibnamefont {Kish}}, \bibinfo {author} {\bibfnamefont {A.~T.}\ \bibnamefont {Sexton}}, \bibinfo {author} {\bibfnamefont {E.~S.}\ \bibnamefont {Evans}}, \bibinfo {author} {\bibfnamefont {I.}~\bibnamefont {Mohammad}}, \bibinfo {author} {\bibfnamefont {S.}~\bibnamefont {Gomez-Ramirez}}, \bibinfo {author} {\bibfnamefont {W.}~\bibnamefont {Scullin}}, \bibinfo {author} {\bibfnamefont {M.}~\bibnamefont {Borscz}}, \bibinfo {author} {\bibfnamefont {S.}~\bibnamefont {Pikuz}}, \bibinfo {author} {\bibfnamefont {T.~A.}\ \bibnamefont {Mehlhorn}}, \emph {et~al.},\ }\bibfield  {title} {\bibinfo {title} {A kinetic study of fusion burn waves in compressed deuterium--tritium and proton--boron plasmas},\ }\href@noop {} {\bibfield  {journal} {\bibinfo  {journal} {Frontiers in Physics}\ }\textbf {\bibinfo {volume} {12}},\ \bibinfo {pages} {1440037} (\bibinfo {year} {2024})}\BibitemShut {NoStop}%
\bibitem [{\citenamefont {Arber}\ \emph {et~al.}(2015)\citenamefont {Arber}, \citenamefont {Bennett}, \citenamefont {Brady}, \citenamefont {Lawrence-Douglas}, \citenamefont {Ramsay}, \citenamefont {Sircombe}, \citenamefont {Gillies}, \citenamefont {Evans}, \citenamefont {Schmitz}, \citenamefont {Bell},\ and\ \citenamefont {Ridgers}}]{EPOCH}%
  \BibitemOpen
  \bibfield  {author} {\bibinfo {author} {\bibfnamefont {T.~D.}\ \bibnamefont {Arber}}, \bibinfo {author} {\bibfnamefont {K.}~\bibnamefont {Bennett}}, \bibinfo {author} {\bibfnamefont {C.~S.}\ \bibnamefont {Brady}}, \bibinfo {author} {\bibfnamefont {A.}~\bibnamefont {Lawrence-Douglas}}, \bibinfo {author} {\bibfnamefont {M.~G.}\ \bibnamefont {Ramsay}}, \bibinfo {author} {\bibfnamefont {N.~J.}\ \bibnamefont {Sircombe}}, \bibinfo {author} {\bibfnamefont {P.}~\bibnamefont {Gillies}}, \bibinfo {author} {\bibfnamefont {R.~G.}\ \bibnamefont {Evans}}, \bibinfo {author} {\bibfnamefont {H.}~\bibnamefont {Schmitz}}, \bibinfo {author} {\bibfnamefont {A.~R.}\ \bibnamefont {Bell}},\ and\ \bibinfo {author} {\bibfnamefont {C.~P.}\ \bibnamefont {Ridgers}},\ }\bibfield  {title} {\bibinfo {title} {{Contemporary particle-in-cell approach to laser-plasma modelling}},\ }\href@noop {} {\bibfield  {journal} {\bibinfo  {journal} {Plasma Physics and Controlled Fusion}\ }\textbf {\bibinfo {volume} {57}},\ \bibinfo {pages} {1} (\bibinfo
  {year} {2015})}\BibitemShut {NoStop}%
\bibitem [{\citenamefont {Haus}\ and\ \citenamefont {Huang}(1991)}]{Haus:1991}%
  \BibitemOpen
  \bibfield  {author} {\bibinfo {author} {\bibfnamefont {H.~A.}\ \bibnamefont {Haus}}\ and\ \bibinfo {author} {\bibfnamefont {W.}~\bibnamefont {Huang}},\ }\bibfield  {title} {\bibinfo {title} {Coupled-mode theory},\ }\href@noop {} {\bibfield  {journal} {\bibinfo  {journal} {Proceedings of the IEEE}\ }\textbf {\bibinfo {volume} {79}},\ \bibinfo {pages} {1505} (\bibinfo {year} {1991})}\BibitemShut {NoStop}%
\bibitem [{\citenamefont {Chen}(2015)}]{FChen:2015}%
  \BibitemOpen
  \bibfield  {author} {\bibinfo {author} {\bibfnamefont {F.}~\bibnamefont {Chen}},\ }\href@noop {} {\emph {\bibinfo {title} {Introduction to plasma physics and controlled fusion}}}\ (\bibinfo  {publisher} {Springer},\ \bibinfo {year} {2015})\BibitemShut {NoStop}%
\bibitem [{\citenamefont {Naumova}\ \emph {et~al.}(2001)\citenamefont {Naumova}, \citenamefont {Bulanov}, \citenamefont {Esirkepov}, \citenamefont {Farina}, \citenamefont {Nishihara}, \citenamefont {Pegoraro}, \citenamefont {Ruhl},\ and\ \citenamefont {Sakharov}}]{Naumova:2001}%
  \BibitemOpen
  \bibfield  {author} {\bibinfo {author} {\bibfnamefont {N.}~\bibnamefont {Naumova}}, \bibinfo {author} {\bibfnamefont {S.}~\bibnamefont {Bulanov}}, \bibinfo {author} {\bibfnamefont {T.~Z.}\ \bibnamefont {Esirkepov}}, \bibinfo {author} {\bibfnamefont {D.}~\bibnamefont {Farina}}, \bibinfo {author} {\bibfnamefont {K.}~\bibnamefont {Nishihara}}, \bibinfo {author} {\bibfnamefont {F.}~\bibnamefont {Pegoraro}}, \bibinfo {author} {\bibfnamefont {H.}~\bibnamefont {Ruhl}},\ and\ \bibinfo {author} {\bibfnamefont {A.}~\bibnamefont {Sakharov}},\ }\bibfield  {title} {\bibinfo {title} {Formation of electromagnetic postsolitons in plasmas},\ }\href@noop {} {\bibfield  {journal} {\bibinfo  {journal} {Physical Review Letters}\ }\textbf {\bibinfo {volume} {87}},\ \bibinfo {pages} {185004} (\bibinfo {year} {2001})}\BibitemShut {NoStop}%
\bibitem [{\citenamefont {Esirkepov}\ \emph {et~al.}(2002)\citenamefont {Esirkepov}, \citenamefont {Nishihara}, \citenamefont {Bulanov},\ and\ \citenamefont {Pegoraro}}]{Esirkepov:2002}%
  \BibitemOpen
  \bibfield  {author} {\bibinfo {author} {\bibfnamefont {T.}~\bibnamefont {Esirkepov}}, \bibinfo {author} {\bibfnamefont {K.}~\bibnamefont {Nishihara}}, \bibinfo {author} {\bibfnamefont {S.~V.}\ \bibnamefont {Bulanov}},\ and\ \bibinfo {author} {\bibfnamefont {F.}~\bibnamefont {Pegoraro}},\ }\bibfield  {title} {\bibinfo {title} {Three-dimensional relativistic electromagnetic subcycle solitons},\ }\href@noop {} {\bibfield  {journal} {\bibinfo  {journal} {Physical review letters}\ }\textbf {\bibinfo {volume} {89}},\ \bibinfo {pages} {275002} (\bibinfo {year} {2002})}\BibitemShut {NoStop}%
\bibitem [{\citenamefont {Lindsey}\ \emph {et~al.}(2024)\citenamefont {Lindsey}, \citenamefont {Bekx}, \citenamefont {Schlesinger},\ and\ \citenamefont {Glenzer}}]{Lindsey:2024}%
  \BibitemOpen
  \bibfield  {author} {\bibinfo {author} {\bibfnamefont {M.~L.}\ \bibnamefont {Lindsey}}, \bibinfo {author} {\bibfnamefont {J.~J.}\ \bibnamefont {Bekx}}, \bibinfo {author} {\bibfnamefont {K.-G.}\ \bibnamefont {Schlesinger}},\ and\ \bibinfo {author} {\bibfnamefont {S.~H.}\ \bibnamefont {Glenzer}},\ }\bibfield  {title} {\bibinfo {title} {Dynamically assisted nuclear fusion in the strong-field regime},\ }\href@noop {} {\bibfield  {journal} {\bibinfo  {journal} {Physical Review C}\ }\textbf {\bibinfo {volume} {109}},\ \bibinfo {pages} {044605} (\bibinfo {year} {2024})}\BibitemShut {NoStop}%
\bibitem [{\citenamefont {Lv}\ \emph {et~al.}(2022)\citenamefont {Lv}, \citenamefont {Duan},\ and\ \citenamefont {Liu}}]{Lv:2022}%
  \BibitemOpen
  \bibfield  {author} {\bibinfo {author} {\bibfnamefont {W.}~\bibnamefont {Lv}}, \bibinfo {author} {\bibfnamefont {H.}~\bibnamefont {Duan}},\ and\ \bibinfo {author} {\bibfnamefont {J.}~\bibnamefont {Liu}},\ }\bibfield  {title} {\bibinfo {title} {Enhanced proton-boron nuclear fusion cross sections in intense high-frequency laser fields},\ }\href@noop {} {\bibfield  {journal} {\bibinfo  {journal} {Nuclear Physics A}\ }\textbf {\bibinfo {volume} {1025}},\ \bibinfo {pages} {122490} (\bibinfo {year} {2022})}\BibitemShut {NoStop}%
\bibitem [{\citenamefont {Qi}\ \emph {et~al.}(2026)\citenamefont {Qi}, \citenamefont {Zhou},\ and\ \citenamefont {Wang}}]{Qi:2026}%
  \BibitemOpen
  \bibfield  {author} {\bibinfo {author} {\bibfnamefont {J.-T.}\ \bibnamefont {Qi}}, \bibinfo {author} {\bibfnamefont {Z.-Y.}\ \bibnamefont {Zhou}},\ and\ \bibinfo {author} {\bibfnamefont {X.}~\bibnamefont {Wang}},\ }\bibfield  {title} {\bibinfo {title} {Theory of laser-assisted nuclear fusion},\ }\href@noop {} {\bibfield  {journal} {\bibinfo  {journal} {Nuclear Science and Techniques}\ }\textbf {\bibinfo {volume} {37}},\ \bibinfo {pages} {53} (\bibinfo {year} {2026})}\BibitemShut {NoStop}%
\bibitem [{\citenamefont {Chowdhury}\ \emph {et~al.}(2019)\citenamefont {Chowdhury}, \citenamefont {Manna}, \citenamefont {Chatterjee}, \citenamefont {Sen},\ and\ \citenamefont {Pal}}]{Chowdhury:2019}%
  \BibitemOpen
  \bibfield  {author} {\bibinfo {author} {\bibfnamefont {S.~D.}\ \bibnamefont {Chowdhury}}, \bibinfo {author} {\bibfnamefont {S.}~\bibnamefont {Manna}}, \bibinfo {author} {\bibfnamefont {S.}~\bibnamefont {Chatterjee}}, \bibinfo {author} {\bibfnamefont {R.}~\bibnamefont {Sen}},\ and\ \bibinfo {author} {\bibfnamefont {M.}~\bibnamefont {Pal}},\ }\bibfield  {title} {\bibinfo {title} {Mega-hertz repetition rate broadband nano-second pulses from an actively mode-locked yb-fiber laser},\ }\href {https://doi.org/10.1088/1555-6611/aafd29} {\bibfield  {journal} {\bibinfo  {journal} {Laser Physics}\ }\textbf {\bibinfo {volume} {29}},\ \bibinfo {pages} {035102} (\bibinfo {year} {2019})}\BibitemShut {NoStop}%
\bibitem [{\citenamefont {Mourou}\ \emph {et~al.}(2013)\citenamefont {Mourou}, \citenamefont {Brocklesby}, \citenamefont {Tajima},\ and\ \citenamefont {Limpert}}]{Mourou:2013}%
  \BibitemOpen
  \bibfield  {author} {\bibinfo {author} {\bibfnamefont {G.}~\bibnamefont {Mourou}}, \bibinfo {author} {\bibfnamefont {B.}~\bibnamefont {Brocklesby}}, \bibinfo {author} {\bibfnamefont {T.}~\bibnamefont {Tajima}},\ and\ \bibinfo {author} {\bibfnamefont {J.}~\bibnamefont {Limpert}},\ }\bibfield  {title} {\bibinfo {title} {The future is fibre accelerators},\ }\href {https://doi.org/10.1038/nphoton.2013.75} {\bibfield  {journal} {\bibinfo  {journal} {Nature Photonics}\ }\textbf {\bibinfo {volume} {7}},\ \bibinfo {pages} {258} (\bibinfo {year} {2013})}\BibitemShut {NoStop}%
\end{thebibliography}%
\end{document}